\documentclass[
   aps,
    prx,
    twocolumn,
    10pt,
    a4paper,
    superscriptaddress,
    notitlepage,
    longbibliography,
    floatfix
]{revtex4-2}
\usepackage[utf8]{inputenc}
\usepackage[english]{babel}
\usepackage{dirtytalk}
\usepackage{amsfonts}
\usepackage{amsmath}
\usepackage{amssymb}
\usepackage{bbold}
\usepackage[dvipsnames,svgnames,table]{xcolor}
\usepackage{graphicx}
\usepackage{fancyhdr}
\usepackage{hyperref}

\usepackage{float}
\usepackage{braket}
\usepackage{float}
\usepackage[caption=false]{subfig}
\usepackage{outlines}

\usepackage{orcidlink}

\hypersetup{colorlinks=true, citecolor=blue, urlcolor=blue, linkcolor=blue}

\usepackage{silence}
\usepackage{tikz}

\hypersetup{
    pdfstartview={FitH},% fits the width of the page to the window
    colorlinks=true,    % false: boxed links; true: colored links
    linkcolor=NavyBlue, % color of internal links
    citecolor=Maroon,   % color of links to bibliography
    filecolor=NavyBlue, % color of file links
    urlcolor=NavyBlue   % color of external links
}

\makeatletter
\def\footnoterule{\kern -10pt
    \hrule \@width 100pt \kern 10pt} % the \hrule is .4pt high
\makeatother

\makeatletter
    \def\bbl@set@language#1{%
      \edef\languagename{%
        \ifnum\escapechar=\expandafter`\string#1\@empty
        \else\string#1\@empty\fi}%
      \@ifundefined{babel@language@alias@\languagename}{}{%
        \edef\languagename{\@nameuse{babel@language@alias@\languagename}}%
      }%
      \select@language{\languagename}%
      \expandafter\ifx\csname date\languagename\endcsname\relax\else
        \if@filesw
          \protected@write\@auxout{}{\string\select@language{\languagename}}%
          \bbl@for\bbl@tempa\BabelContentsFiles{%
            \addtocontents{\bbl@tempa}{\xstring\select@language{\languagename}}}%
          \bbl@usehooks{write}{}%
        \fi
      \fi}
    \newcommand{\DeclareLanguageAlias}[2]{%
      \global\@namedef{babel@language@alias@#1}{#2}%
    }
\makeatother
\DeclareLanguageAlias{en}{english}

\newcommand{\KVSO}{KV$_{2}$Se$_{2}$O}

\newcommand{\ucr}{UCr$_2$Si$_2$C}
\newcommand{\loas}{La$_2$O$_3$A$_2$Se$_2$}
\newcommand{\vso}{V$_2$Se$_2$O}
\newcommand{\rbvto}{RbV$_2$Te$_2$O}
\newcommand{\srcroas}{Sr$_2$CrO$_2$Cr$_2$OAs$_2$}

\bibpunct{[}{]}{,}{n}{}{}  % show bibliography in-line.

\begin{document}

    \title{Emergence and Detection of Surface altermagnetism in \texorpdfstring{\KVSO{}}{KV2Se2O}}
        
\author{Rodrigo Jaeschke-Ubiergo  \orcidlink{0000-0002-4821-8303}}
\thanks{These authors contributed equally to this work}
\affiliation{Institute of Physics, Johannes Gutenberg University Mainz, 55099 Mainz, Germany}

    \author{Xanthe H. Verbeek  \orcidlink{0000-0003-2262-5047} $^*$}
        %\email{xverbeek@uni-mainz.de}
        \thanks{These authors contributed equally to this work}
       \affiliation{Institute of Physics, Johannes Gutenberg University Mainz, 55099 Mainz, Germany}

   \author{Colin Lange \orcidlink{0009-0001-0661-7998}}
\affiliation{Institute of Physics, Johannes Gutenberg University Mainz, 55099 Mainz, Germany}

    \author{Sergio Rodríguez Fernández}
        \affiliation{Institute of Physics, Johannes Gutenberg University Mainz, 55099 Mainz, Germany}

\author{Atasi Chakraborty}
\affiliation{Institute of Physics, Johannes Gutenberg University Mainz, 55099 Mainz, Germany}
\author{Alexander Mook \orcidlink{0000-0002-8599-9209}}
\affiliation{University of M\"{u}nster, Institute of Solid State Theory, 48149 Münster, Germany}
   \author{Jairo Sinova \orcidlink{0000-0002-9490-2333}}
\affiliation{Institute of Physics, Johannes Gutenberg University Mainz, 55099 Mainz, Germany}
\affiliation{Department of Physics, Texas A\&M University, College Station, Texas 77843-4242, USA}

\begin{abstract}
We demonstrate the recent concept of emergent surface altermagnetism through its 
unique signatures  in \KVSO. We show that for bulk antiferromagnetically ordered \KVSO, the (001) surface exhibits $d$-wave altermagnetism. Our results fully explain the recent seemingly contradicgting experimental evidence, independently showing both an antiferromagnetically ordered bulk from neutron diffraction, and $d$-wave spin splitting from photoemission spectroscopy. To fully verify this conecept, we predict, as a key experimental signature,  a large nonlinear Edelstein response, which is localized at the surface, and follows the $d$-wave altermagnetic symmetry. These results are not only relevant for the metallic and room-temperature magnet \KVSO, but also for several other Lieb lattice systems. Consequently, our work expands the pool of techniques that can be used to detect altermagnetism emerging at the surfaces of antiferromagnets.
\end{abstract}

\maketitle

\begin{figure*}[t!]
  \centering
     \includegraphics[width=\linewidth]{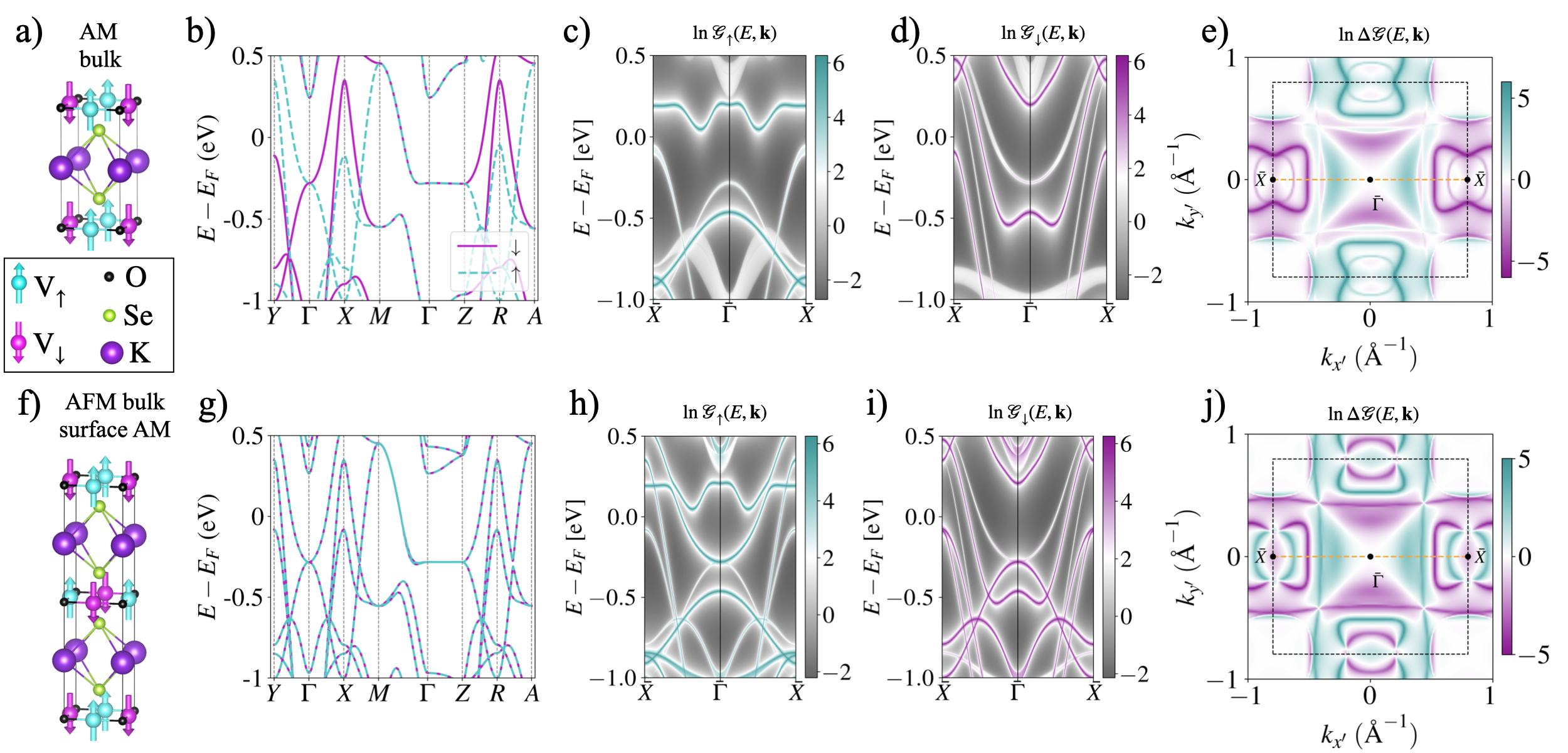}
    \caption{Non-relativistic DFT calculations of \KVSO\ with AM (a-e) and AFM (f-j) ordering. (a) Unit cell of the AM configuration, cyan and magenta arrows denote magnetic moments on the V atoms with opposite direction. (b) Spin-polarized bulk electronic band structure, with altermagnetic $d$-wave splitting. (c-e) Surface states for the VO-terminated $(001)$ surface of the AM configuration. (c) Spin-up and (d) spin-down channel surface spectral function ($\mathcal{G}_{\sigma}(E,\mathbf{k})$), in a spin-split path. (e) The spin-polarized spectral function difference ($\Delta \mathcal{G}(E,\mathbf{k})=\mathcal{G}_{\uparrow}(E,\mathbf{k})-\mathcal{G}_{\downarrow}(E,\mathbf{k})$) at the Fermi energy $E_{F}$ exhibiting $d$-wave character with two nodal planes. Surface Brillouin zone indicated with a black dashed square. (f) Magnetic unit cell of the AFM configuration. (g) Kramers degenerate bulk band structure. (h-j) Surface states calculations for the VO-terminated $(001)$ surface of the AFM configuration. (h) Spin-up and (i) spin-down channel spectral function ($\mathcal{G}_{\sigma}(E,\mathbf{k})$) in a spin-split path. (j) The spin-polarized spectral function difference ($\Delta \mathcal{G}(E,\mathbf{k})=\mathcal{G}_{\uparrow}(E,\mathbf{k})-\mathcal{G}_{\downarrow}(E,\mathbf{k})$) at the Fermi energy $E_{F}$.}
    \label{fig:KVSO_AM_vs_AFM}
\end{figure*}

\section{Introduction}
\label{sec::introduction}

Altermagnets (AMs) are collinear and compensated magnets in which the electronic band structure displays exchange-driven spin-splitting, with alternating sign in momentum space and characteristic $d$, $g$ or $i$-wave symmetry \cite{Smejkal2021a}.
Due to their abundance and unconventional electronic band structure that breaks time-reversal symmetry despite their vanishing net magnetization \cite{Smejkal2022a,Jungwirth2025}, altermagnets offer a plethora of effects \cite{Smejkal2020,Mazin2021,Ahn2019,Naka2019,Gonzalez-Hernandez2021,Smejkal2022GMR,Samanta2020,Mazin2021,Smejkal2020,Smejkal2022GMR,Bai2024,Naka2019,Shao2021,Zhu2023a,Jaeschke-Ubiergo2023,Karetta2025,Jaeschke-Ubiergo2025,Trama2025,Golub2025} that are rapidly paving the road for energy-efficient next-generation spintronic technologies \cite{Smejkal2022a,Jungwirth2025b,Song2025}.

Since their original prediction \cite{Smejkal2021a,Smejkal2022a}, which was based on a systematic symmetry classification of all collinear magnetic phases, there has been increasing experimental evidence of the altermagnetic phase. This evidence goes from direct observation of the spin splitting by photoemission spectroscopy in the room temperature compounds MnTe and CrSb \cite{Krempasky2024,Lee2024,Osumi2024,Reimers2024,Yang2024,Ding2024,Zeng2024,Li2024,Hariki2023,Amin2024}, to indirect probes of both the time-reversal-symmetry breaking, and the spin order with even-parity-wave anisotropy. The latter include the anomalous Hall and Nernst effects \cite{Feng2022,Betancourt2021,Reichlova2024,Leiviska2024,Han2024,Galindez-Ruales2025a,Badura2025,Mali2026}, unconventional spin-polarized and spin-splitter currents \cite{Bose2022,Bai2022,Karube2022,Liu2023,Liao2024}, magneto-optical effects \cite{Hariki2023, Amin2024,Fedchenko2024}, and piezo-magnetism \cite{Aoyama2024}, among others. The research landscape in altermagnetism is not bounded to electronic systems. Magnetic excitations in altermagnets \cite{Smejkal2023} inherit the even-parity wave spin splitting \cite{Smejkal2023, Naka2019, Gohlke2022, Gomonay2024, Liu2024b,Sun2025b}, and can host unconventional thermal transport responses \cite{Hoyer2025,Hoyer2025a,Galindez-Ruales2025,Weissenhofer2024}.
Two-dimensional altermagnets \cite{Mazin2023a,Zeng2024a,Jana2026}, are of particular interest because many applications rely on interface effects, and two-dimensional systems provide a natural playground for building layered heterostructures. 

Recently, the prediction of surface altermagnetism \cite{Lange2026} has considerably expanded the list of altermagnetic candidates. The central idea is that certain surfaces of some antiferromagnets will lower their bulk symmetry, breaking time-reversal symmetry, while keeping a compensating symmetry that enforces the alternating spin polarization of the electronic surface states.
Surface altermagnetism is becoming relevant \cite{Lange2026,Zhou2025,Leeb2026,Belashchenko2026,Franz2026,Sasioglu2026} because it effectively expands the number of altermagnets to be used as active elements in two dimensions, and also because ARPES, one of the most direct techniques to measure the altermagnetic splitting, is intrinsically a probe of the surface. A key example in this regard is the evidence of $d$-wave altermagnetism in the metallic room-temperature \KVSO\ \cite{Jiang2025}, which was in apparent conflict with a more recent neutron diffraction experiment \cite{Sun2025a}, which measured a layered antiferromagnetic ordering with propagation vector $(0,0,\frac{1}{2})$, which makes \KVSO\ in principle incompatible with altermagnetism. This material and other layered systems with Lieb lattice geometry \cite{Lin2024,Qi2024,Zhang2025a,Wei2025,Jaeschke-Ubiergo2025,Campos2026}, are some of the numerous examples of surface altermagnetism \cite{Lange2026}. At the $(001)$ surface, $\mathcal{T}\vec{t}$ and $\mathcal{P}\mathcal{T}$ are broken, with $\mathcal{T}$, $\mathcal{P}$ and $\vec{t}$ denoting time-reversal symmetry, inversion symmetry and a half-lattice translation, respectively. 
However, there is still a set of spin symmetries that connect magnetic moments with opposite directions on each layer independently. An example of such a symmetry is $[C_2|| C_{4z}]$, which combines a four-fold rotation around $z$-axis in real space with a two-fold rotation in spin space. The presence of these symmetries enforces the magnetic compensation and the $d$-wave spin splitting at the surface.

In this work, we present a detailed analysis of both the altermagnetic and antiferromagnetic phases of \KVSO. Using Density Functional Theory (DFT), we study the bulk electronic structure and the corresponding surface states. We build on the work of Ref.~\cite{Lange2026}, which showed how the antiferromagnetic phase is expected to show $d$-wave altermagnetic splitting at the $(001)$ surface. We extend these results by showing how different surface terminations affect the surface states. Surface altermagnetism in \KVSO{} would explain why neutron diffraction experiments measure an antiferromagnetic ordering in the bulk \cite{Sun2025a}, while spectroscopy measurements \cite{Jiang2025} and spin-selective scanning tunneling microscopy \cite{Yang2026} give evidence of altermagnetic splitting. 

We also propose the nonlinear Edelstein effect as a key experimental signature of the surface altermagnetism in \KVSO. The Edelstein effect refers to the non-equilibrium spin density induced by an applied electric field \cite{Edelstein1990}. The linear Edelstein response is usually encountered in non-centrosymmetric systems with strong spin-orbit coupling (SOC), while its origin can be non-relativistic in systems with odd-parity-wave magnetism \cite{Hellenes2023a,Chakraborty2024c,Pari2025}. Recently, the nonlinear contribution to the Edelstein effect \cite{Xu2021,Baek2024}, was predicted to be present in $d$-wave altermagnets \cite{Trama2025,Golub2025}. This effect usually arises from the interplay of time-reversal symmetry breaking and spin-orbit coupling.

Using Wannier functions, we build a finite slab of the antiferromagnetic \KVSO{}, and we calculate the nonlinear Edelstein response. By resolving the response on the layers, we show that the effect cancels out in the bulk, and peaks at the altermagnetic surface. Notably, the symmetries of \KVSO{} allow for complete disentanglement of the linear and nonlinear contributions to the Edelstein response. The latter one, which is connected to the surface altermagnetism, induces an out-of-plane spin density, while the linear response is in-plane. For realistic values of the electric field, the nonlinear spin density is comparable to or even larger than the linear contribution.

This article is organized as follows. We first show in Sec.\ref{sec:surface_am}, a comparison of the electronic band structures of both the altermagnetic and antiferromagnetic phases of \KVSO. We see how both phases show $d$-wave splitting of the surface states at the $(001)$ surface, even though the bulk AFM is completely spin degenerate due to $\mathcal{PT}$ symmetry. We then introduce in Sec.~\ref{sec:surface_nlee} a finite slab of \KVSO{}, and we discuss the nonlinear Edelstein response (Sec~\ref{sec:nlee_intro}). In Sec.~\ref{sec:nlee_lr}, we explore different surface terminations and the dependence on the number of layers. Finally, in Sec.~\ref{sec:nlee_comparison_lee}, we compare the nonlinear response with the linear counterpart, both in symmetry and magnitude.

\begin{figure*}[t]
  \centering
     \includegraphics[width=\textwidth]{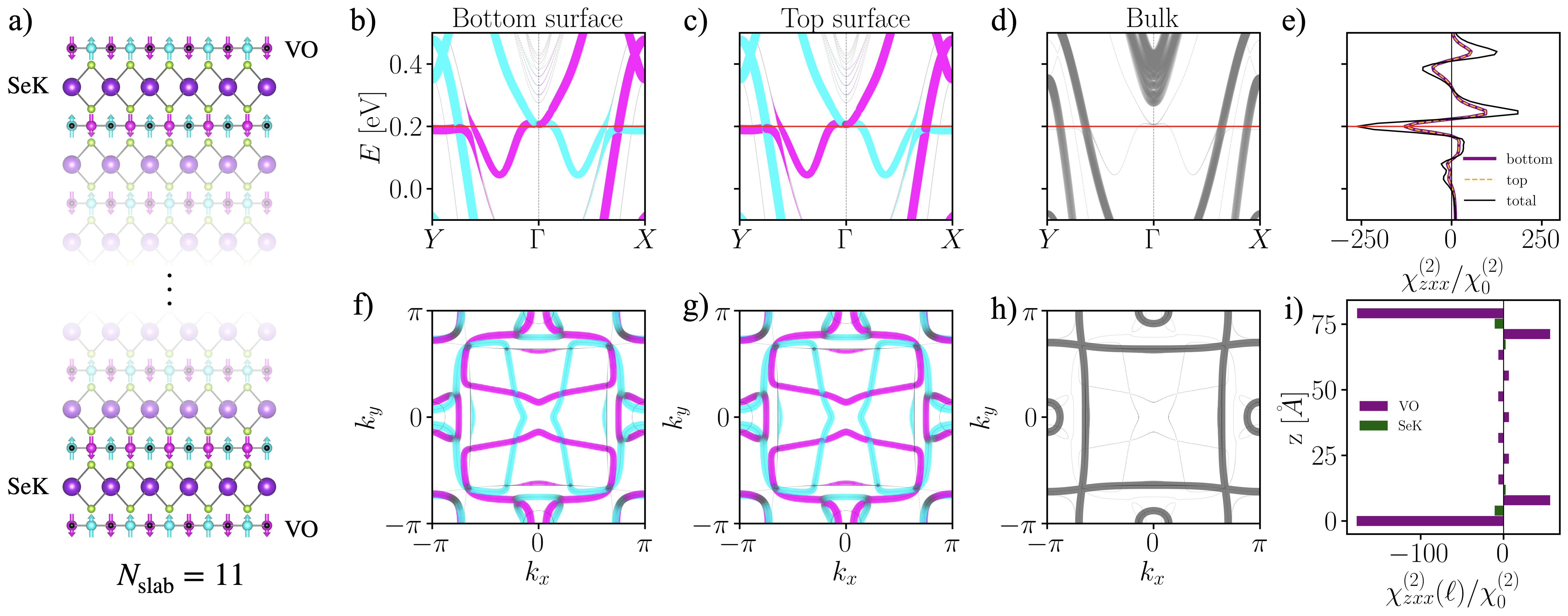}
    \caption{Calculations including SOC on a slab of \KVSO{}. (a) Slab of \KVSO, with $N_{\text{slab}}=11$ and broken $\mathcal{PT}$. (b-d) Spin-polarized band structure projected on the bottom surface (b), top surface (c) and bulk (d). (f-h) Spin-polarized isoenergy lines at $\mu=E_F+ 0.2 \rm eV$, projected on the bottom surface (g), top surface (g) and bulk (h). (e) $\chi_{zxx}$ component of the NLEE tensor, as function of the chemical potential, resolved in both surfaces. (i)Layer-resolved NLEE tensor at $\mu= 0.2 \rm eV$. We define $\chi^{(2)}_0=\frac{e^2 \tau^2 a^2}{(2\pi)^2 \hbar}$.}
    \label{fig:KVSO_slab_nlee}
\end{figure*}

\section{Surface Altermagnetism in KV$_2$Se$_2$O}
\label{sec:surface_am}

We start by comparing the antiferromagnetic and altermagnetic configurations of \KVSO{}. We examine both the electronic band structure in the bulk and the surface states on the (001) surface, the latter of which was probed experimentally \cite{Jiang2025, Yang2026}. \KVSO\ crystallizes in a tetragonal structure with space group $P4/mmm$ (No.123), with the O atoms sitting on Wyckoff position 1a~$(0,0,0)$, the K atoms on 1b~$(0,0,\frac{1}{2})$, the V atoms on 2f~$(\frac{1}{2},0,0)$, and the Se atoms on position 2h~$(\frac{1}{2}, \frac{1}{2}, z)$. 

In the AM bulk configuration, with magnetic space group $P4'/mm'm$ (BNS 123.342), the magnetic and structural unit cells are equivalent (see Fig.~\ref{fig:KVSO_AM_vs_AFM}a), with a plane of V and O atoms forming a Lieb lattice, separated by a layer of K and Se atoms, with the layers stacked along the c axis. Its spin point group ${}^24/{}^1m{}^1m{}^2m$ classifies it as a $d$-wave altermagnet, with nodal planes $(110)$ and $(1\bar{1}0)$. The non-relativistic spin-split band structure, obtained from our ab initio calculations, is shown in Fig.~\ref{fig:KVSO_AM_vs_AFM}b. Since the system is already altermagnetic in the bulk, the corresponding surface states show the altermagnetic splitting as well. We calculate the surface spectral function a for the spin channel $\sigma$ as $\mathcal{G}_{\sigma}(E,\mathbf{k})=-\frac{1}{\pi} \mathrm{Im} \mathrm{Tr} \;G_{\sigma}(E, \mathbf{k})$, with $G_{\sigma}(E,\mathbf{k})$ denoting the surface Green's function of the spin channel $\sigma$, which is calculated using the surface Green's function renormalization formalism as implemented in the WannierTools package \cite{Wu2017b} (see computational details in the Appendix~\ref{sec:comp-details}). Figures~\ref{fig:KVSO_AM_vs_AFM}c,d show the surface spectral function $\mathcal{G}_{\sigma}(E, \mathbf{k})$ for both spin channels $\sigma \in \{\uparrow, \downarrow \}$. Figure~\ref{fig:KVSO_AM_vs_AFM}e shows the spin-polarized surface spectral function across the Brillouin Zone at the Fermi energy. In this calculation, we have picked a surface terminated in the VO layer. 

In contrast, for the AFM ordering proposed by the neutron diffraction experiment of Ref.~\cite{Sun2025a}, the system orders with a propagation vector $(0,0,\frac{1}{2})$, and magnetic space group $P_c 4_2 / mcm$ (BNS 132.456). The structural unit cell is doubled along the $c$-axis, as shown in Fig.~\ref{fig:KVSO_AM_vs_AFM}f. Here, V sites in consecutive VO planes have magnetic moments with opposite orientation, which introduces $\mathcal{T}\tau$ and $\mathcal{PT}$ symmetries. Consequently, the bulk electronic band structure shows Kramers' spin degeneracy (see Fig. \ref{fig:KVSO_AM_vs_AFM}g).
Despite being antiferromagnetic in the bulk, this configuration is predicted to have surface altermagnetism at the $(001)$ surface \cite{Lange2026}. Figures~\ref{fig:KVSO_AM_vs_AFM}h-j show the spin-split surface states, with the $d$-wave character consistent with the surface spin point group ${}^24{}^2m{}^1m$.
There are strong similarities between the surface-projected spectral function in the AM and the AFM orderings. This may help to reconcile the apparent contradiction between the recent ARPES \cite{Jiang2025} and neutron diffraction \cite{Sun2025a} experiments. If \KVSO{} orders antiferromagnetically with a $(0,0,\frac{1}{2})$ propagation vector, a surface-sensitive probe like ARPES is expected to qualitatively show the same spin-splitting pattern as the bulk altermagnetic configuration. In the next section, we will focus on the surface Edelstein effect, which can be exploited to probe the surface altermagnetism in \KVSO.

\begin{figure}[t]
  \centering
     \includegraphics[width=0.8\columnwidth]{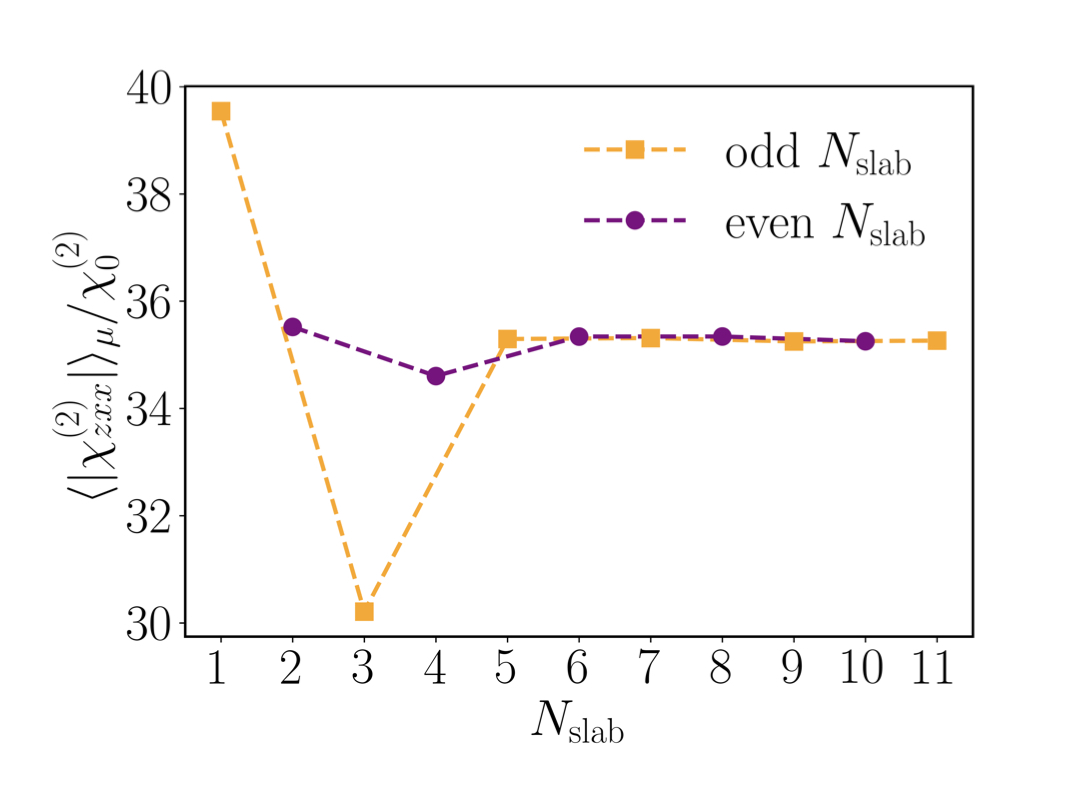}
    \caption{Chemical potential average of the surface nonlinear Edelstein tensor $\frac{1}{\Delta \mu} \int d\mu \; |\chi^{(2)}_{zxx}(\ell=0)|$, over the energy interval $\mu \in [-0.1, 0.5]$ eV, as a function of the number of layers in the slab, for even and odd number of layers.}
    \label{fig:nlee_vs_nslab}
\end{figure}

\begin{figure*}[t]
  \centering
     \includegraphics[width=\textwidth]{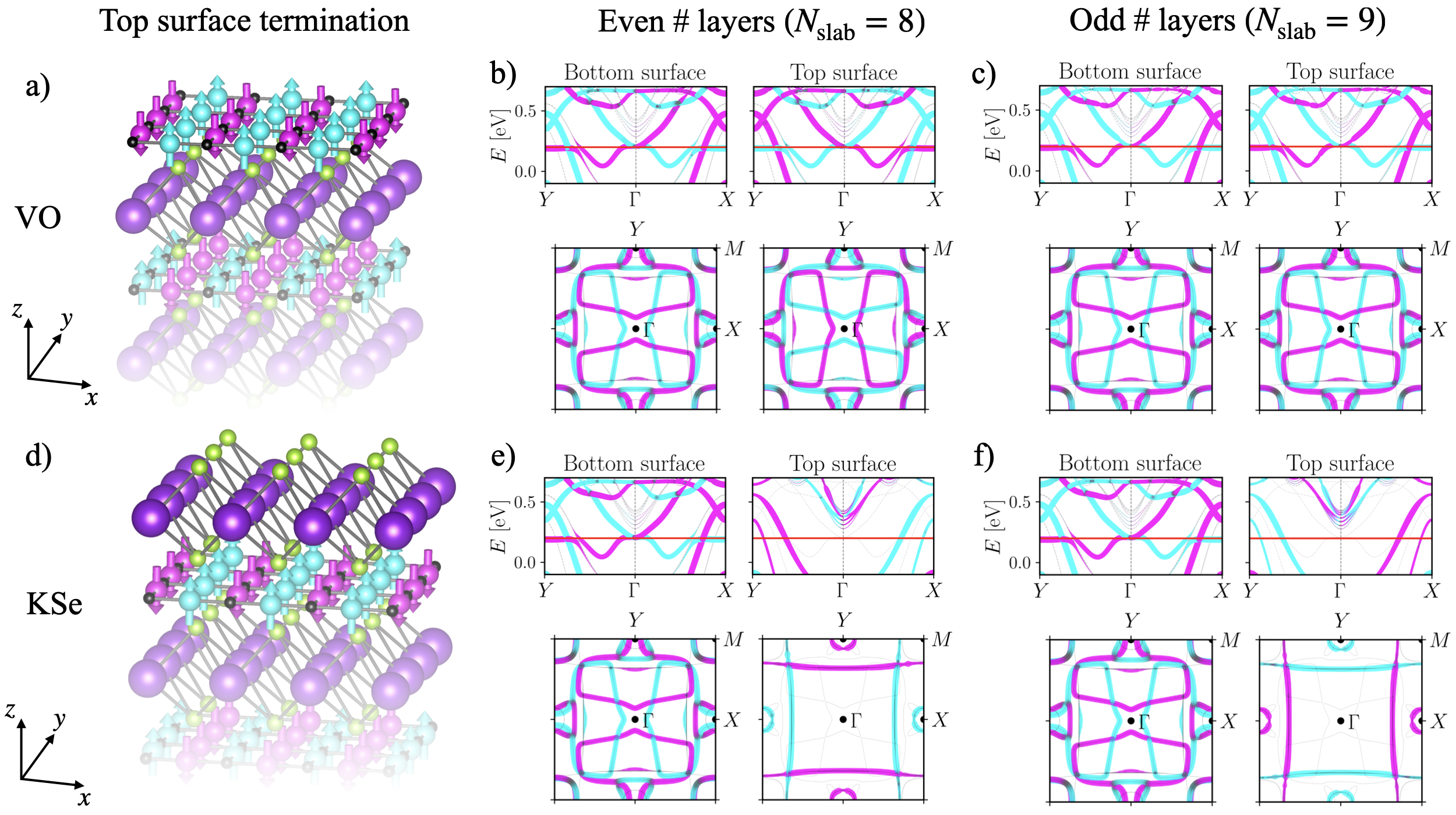}
    \caption{Slab calculations of \KVSO\ (including SOC) with two types of terminations and number of layers. (a,d) Graphic representation of the terminations used in the top surface. (a) Termination in VO plane. (d) Termination in SeK layers. Bottom termination (not shown) is VO in both cases. (b,c) Electronic structure for VO top surface termination with even (b) and odd (c) number of layers. (e,f) Electronic structure for SeK top-surface termination with even (c) and odd (f) number of layers. In each of the four cases (b,c,e,f) the electronic structure is shown in 4 subpanels containing the electronic band structure, projected on the bottom (left) and top (right) surfaces. Below each band structure, we show isoenergy lines at $\mu=0.2$ eV, projected on the bottom (left) and top (right) surfaces. In all plots, the width represents the projection, and the color denotes the $S_z$ spin polarization.}
    \label{fig:KVSO_slab_comparison_terminations}
\end{figure*}

\begin{figure}[t]
  \centering
     \includegraphics[width=0.9\columnwidth]{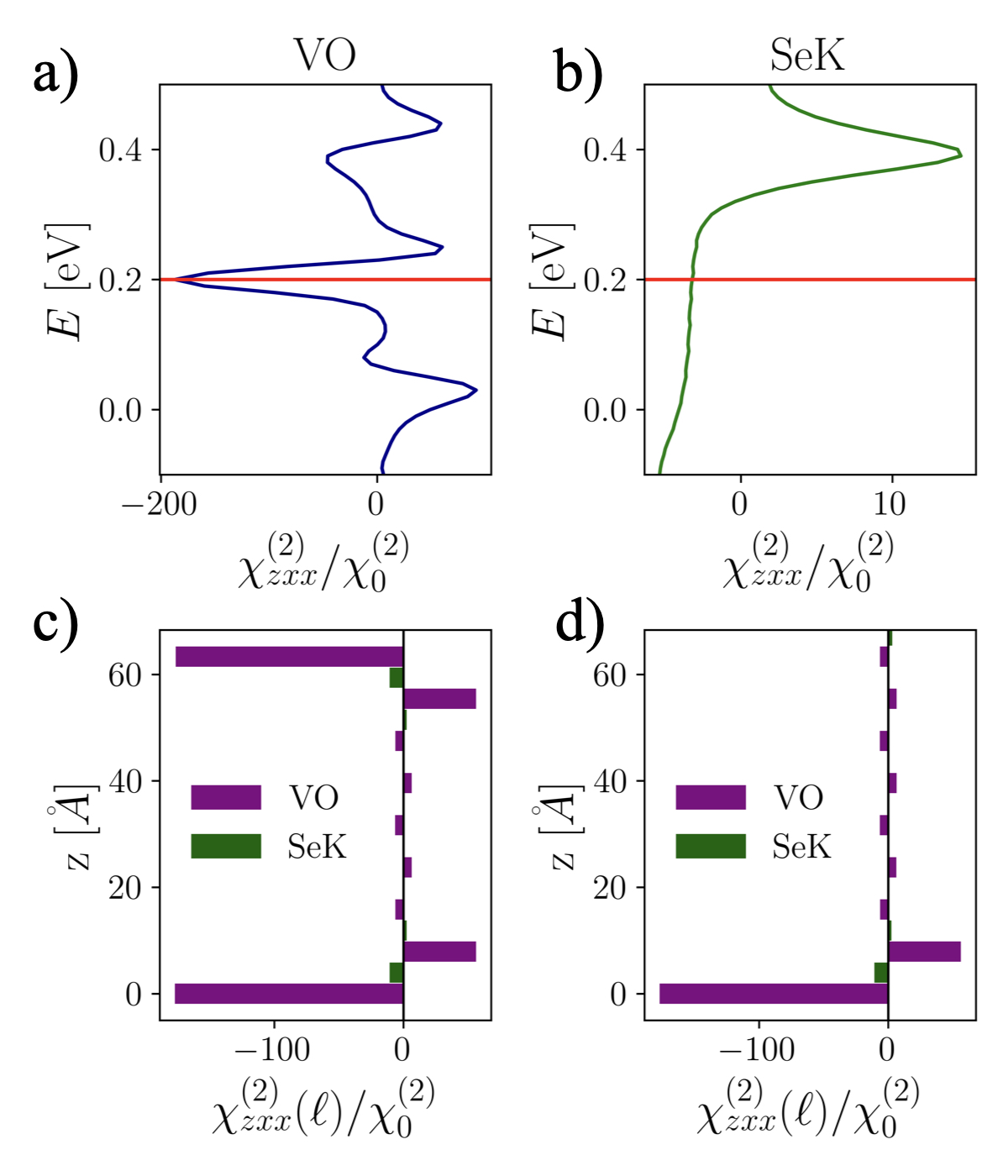}
    \caption{Nonlinear Edelstein response $\chi^{(2)}_{zxx}$ at the top surface of \KVSO{}, with termination on VO (a) and SeK (b). Layer-resolved nonlinear Edelstein response $\chi^{(2)}_{zxx}(\ell)$ for top surface terminated in VO (c) and SeK (d), at $\mu=0.2 \;\mathrm{ eV}$. The bottom surface is terminated at the VO plane in all cases. The slab calculation contains $N_{\mathrm{slab}}=7$ VO planes.}
    \label{fig:nlee__vs_termination}
\end{figure}

\section{Surface nonlinear Edelstein effect}
\label{sec:surface_nlee}

\subsection{Edelstein response and symmetry considerations}
\label{sec:nlee_intro}
Within the Boltzmann transport formalism, one can expand the non-equilibrium spin density $\delta \mathbf{s}$ induced by an electric field, in powers of $\mathbf{E}$ as
\begin{equation}
\delta s^{\alpha}
=
\chi_{\alpha i}^{(1)}\,E_i
+
\chi_{\alpha ij}^{(2)}\,E_i E_j
+O(E^3),
\end{equation}
where $\alpha = x,y,z$, $i = x,y,z$, and $\chi_{\alpha i}^{(1)}$ and $\chi_{\alpha ij}^{(2)}$ denote the linear and quadratic Edelstein susceptibility tensors, respectively.

The bulk structure of antiferromagnetic \KVSO\ is characterized by the magnetic point group $4/mmm 1'$, which has both inversion and time-reversal symmetry. As a result, both the linear and quadratic Edelstein susceptibilities vanish in the bulk. For a surface with normal $(001)$, the symmetry reduces to $4'm'm$. Both time-reversal and inversion symmetry are broken, and one can expect an electric field-induced spin density at both linear and quadratic orders in the field. The surface symmetry group constrains the linear and quadratic susceptibility tensors to have only in-plane and out-of-plane spin components, respectively. At the surface, one expects the spin density to be given by (see Appendix \ref{symmetry_edelstein}).
\begin{equation}
\delta \mathbf{s}
=
\chi^{(1)}_{xy} \hat{\mathbf{z}}\times \mathbf{E}
+
\chi^{(2)}_{zxx} (E_x^2 - E_y^2) \hat{\mathbf{z}},
\label{eq:edelstein_linear_and_nonlinear}
\end{equation}
where $\hat{\mathbf{z}}$ is a unit vector along the $z$-direction ($c$ axis).

The nonlinear contribution has the same $d$-wave symmetry as the spin splitting induced by the surface altermagnetism. If the surface breaks inversion but not time-reversal symmetry, one would expect only a linear contribution to the Edelstein effect, and the nonlinear term would vanish. Due to the spin-component separation in Eq.~\ref{eq:edelstein_linear_and_nonlinear}, the presence of a nonzero $\chi^{(2)}_{zxx}$ is a direct consequence of the time-reversal symmetry-breaking surface altermagnetism.

Within the constant relaxation time approximation, and neglecting interband transitions, the susceptibilities for the linear and nonlinear contributions at zero temperature are given by
\begin{equation}
\chi_{\alpha i}^{(1)} = -e \tau \sum_n \int d \mathbf{k} \;s_{n}^{\alpha}(\mathbf{k}) v^i_{n}(\mathbf{k}) \delta(\varepsilon_n(\mathbf{k})-E_F)
\end{equation}
and 
\begin{equation}
    \chi^{(2)}_{\alpha,ij}
=
\chi^{(2)}_0
\sum_{n}
\int d^2\mathbf{k}\;
\frac{\partial s^{\alpha}_{n}(\mathbf{k})}{\partial k_i}
\,v^{j}_{n}(\mathbf{k})
\,
\delta\!\left(\varepsilon_n(\mathbf{k})-E_F\right) \;,
\label{chi_2}
\end{equation}
respectively. Here,  $s^{\alpha}_n(\mathbf{k}) = \bra{\psi_{n \mathbf{k}}} \frac{\hbar}{2}\sigma_{\alpha}\ket{\psi_{n \mathbf{k}}}$ is the $\alpha$ component of the spin expectation value of the eigenstate $\ket{\psi_{n \mathbf{k}}}$, with crystal momentum $\mathbf{k}$ and band index $n$. $v^{i}_{n\mathbf{k}}= \bra{\psi_{n \mathbf{k}}} v^i \ket{\psi_{n \mathbf{k}}}$ is the expectation value of the component $i$ of the velocity operator $v^i$, and $\chi^{(2)}_0=\frac{e^2 \tau^2 a^2}{(2\pi)^2 \hbar}$, with $\tau$, $e>0$, and $a=3.984$ \AA{} denoting the constant relaxation time, the electron charge and the in-plane lattice constant of \KVSO{}, respectively.

\subsection{Layer-resolved Edelstein response}
\label{sec:nlee_lr}
In order to estimate the nonlinear Edelstein effect coming from the surface altermagnetism, we build a slab system from the bulk Wannier functions of the antiferromagnetic \KVSO. We resolve this quantity on the different layers by projecting the spin operator on a given layer $\ell$ as
\begin{equation}
s^{\alpha}_{n,\ell}(\mathbf{k}) = \bra{\psi_{n\mathbf{k}}}P_{\ell} (\mathbb{1}_{\rm orb}\otimes \frac{\hbar}{2}\sigma_{\alpha}) P_{\ell} \ket{\psi_{n\mathbf{k}}},
\end{equation}
 where $\mathbb{1}_{\rm orb}$ is the identity matrix in the orbital channel, $P_{\ell}=\sum_{i \in \ell} \ket{\phi_{i\mathbf{k}}}\bra{\phi_{i\mathbf{k}}}$ is a projector on layer $\ell$, and  $\ket{\phi_{i\mathbf{k}}}=\frac{1}{\sqrt{N}}\sum_{\mathbf{R}} e^{i\mathbf{k}\cdot (\mathbf{R}+\mathbf{r}_i)} \ket{\phi_i(\mathbf{R})}$ denotes the normalized Bloch summation of the Wannier orbital $\ket{\phi_i(\mathbf{R})}$ at position $\mathbf{r}_i$ with respect to the unit cell origin at $\mathbf{R}$. The Wannier orbitals are labeled with index $i$, which captures both orbital and spin degrees of freedom. In our calculation, the Wannier basis includes orbitals V-d, Se-p, K-p and O-p. The summation inside each projector $P_{\ell}$ includes all orbitals belonging to layer $\ell$. We define two types of layers, the first one by grouping all V and O sites on a single VO plane, and the second one formed by the Se and K sites between VO planes.

 With this projected spin operator, we can write thlayer-resolveded nonlinear Edelstein tensor as
 \begin{equation}
     \chi^{(2)}_{\alpha ij}(\ell) = \chi^{(2)}_0
\sum_{n}
\int d^2\mathbf{k}\;
\frac{\partial s^{\alpha}_{n,\ell}(\mathbf{k})}{\partial k_i}
\,v^{j}_{n}(\mathbf{k})
\,
\delta\!\left(\varepsilon_n(\mathbf{k})-E_F\right) \;.
\label{eq:nlee_lr}
 \end{equation}

In Fig.~\ref{fig:KVSO_slab_nlee}, we show a slab calculation in \KVSO{} with 11 layers and a VO plane as termination for both the top and bottom surfaces. Here, as elsewhere in the document, we use the number of layers to refer to the number of VO planes. Thus, for a double VO plane terminated stack, 11 layers means 11 VO planes and 10 SeK planes.
Figures~\ref{fig:KVSO_slab_nlee}a-c show the band structure, projected onto the surfaces and the bulk. The lines' thickness represents the projected weight, and the color represents the $s^{z}_{n}(\mathbf{k})$ spin polarization. Since we have an odd number of layers, top and bottom surface show identical surface states. Figure~\ref{fig:KVSO_slab_nlee}d shows the nonlinear Edelstein tensor, projected onto the surfaces and bulk. The horizontal red line at $\mu=0.2$ eV highlights a peak in the response, which matches the energy of the surfaces states, whose character is dominated by V-$d$ orbitals at the respective surface. At this particular chemical potential, we show in Fig.~\ref{fig:KVSO_slab_nlee}f-h the isoenergy lines, with varying width representing the projection in the bottom surface, top surface, and bulk, respectively, and the color denoting the spin polarization. In Fig.~\ref{fig:KVSO_slab_nlee}i, we show the layer resolved nonlinear Edelstein response $\chi^{(2)}_{\alpha ij}(\ell)$ at the same  chemical potential ($\mu=0.2$ eV). We see that the response is maximum at the surface, and rapidly decays inside the bulk. In the bulk, the response per layer is nonzero and alternates sign, due to the $(0,0,1/2)$ propagation vector of the antiferromagnetic order. The response is dominated by the V surface states, and it comes mainly from those regions in Figs.~\ref{fig:KVSO_slab_nlee}e-g in which the spin expectation value is not fully polarized along $z$ (i.e., in the vicinity of the $X$ point) because this leads to a large $\frac{\partial s^{\alpha}_{n}(\mathbf{k})}{\partial k_i}$ in Eq.~\ref{eq:nlee_lr}.

Because the magnetic structure of bulk \KVSO\ is antiferromagnetic, there are effects related to the parity of the number of layers $N_\text{slab}$. Figure~\ref{fig:KVSO_slab_nlee} shows that for an odd number of layers, when the magnetic moments of top and bottom surfaces are aligned, the surface nonlinear Edelstein response is identical in both surfaces. On the other hand, if we pick an even number of layers, while keeping the VO termination, the spin polarization of surface states coming from the top and bottom surfaces is opposite. Similarly, the corresponding layer contributions to the nonlinear Edelstein response have equal magnitude and opposite sign. 
Since the effect is dominant at the surface, as long as the sample is thick enough it should be possible to observe only one of them, independently of the parity of $N_\text{slab}$. 

It is worth mentioning that the surface altermagnetism in \KVSO, as pointed out in Ref.~\cite{Lange2026}, is sensitive to terrace effects. If the surface contains terraces in which the V sites at the termination belong to layers with opposite magnetic moments, the signal will be reduced by averaging over these terraces. In order to experimentally observe the effect, the magnetic domains must be sufficiently large such that cancellation between opposite domains is incomplete.
Since in \KVSO{} there are already experimental signatures of surface altermagnetism \cite{Jiang2025, Yang2026}, it is reasonable to think that such terrace domains will be large enough to not cancel the nonlinear Edelstein response locally, as well as any other time-reversal odd responses which may offer insights into the surface altermagnetism,

We also performed calculations while varying the thickness of the system. In Fig.~\ref{fig:nlee_vs_nslab}, we show the chemical potential average $\langle |\chi^{(2)}_{zxx}|\rangle_\mu= \frac{1}{\Delta \mu} \int d\mu |\chi^{(2)}_{zxx}(\mu)|$ of the response on one of the surfaces as a function of $N_\text{slab}$. We see that the response converges to a single value for both even and odd numbers of layers, which justifies the relatively small $N_{slab}=11$ used in Fig.~\ref{fig:KVSO_slab_nlee}.

In addition to the terraces, the surface termination can significantly affect the response. If the top surface is SeK-terminated, the response at $\mu=0.2$ eV is substantially reduced. To explain this behavior, we first analyze in Fig.~\ref{fig:KVSO_slab_comparison_terminations} the electronic structure, projected onto both top and bottom surfaces, of slab systems with different top-surface terminations, and with even and odd numbers of VO layers ($N_{\mathrm{slab}}=8,\;9$).

Figures~\ref{fig:KVSO_slab_comparison_terminations}a and d show structural representations of the top surfaces with VO and SeK terminations, respectively. The bottom surface (not shown) is terminated at VO planes in both cases. Figures~\ref{fig:KVSO_slab_comparison_terminations}b, c, e, and f are composed of four subpanels each, showing surface-projected band structures and isoenergy lines at $\mu=0.2$ eV.

Figures~\ref{fig:KVSO_slab_comparison_terminations}a-c focus on slabs with a VO-terminated top surface. Figures~\ref{fig:KVSO_slab_comparison_terminations}b and c show the electronic structure for even and odd numbers of VO layers, respectively. With an even number of layers ($N_{\mathrm{slab}}=8$), the bands localized at the top and bottom surfaces have identical dispersions and opposite spin expectation values, because they are related by $\mathcal{P}\mathcal{T}$ symmetry, with $\mathcal{P}$ denoting inversion symmetry. This comes from the alternation of the magnetic moments on consecutive VO planes, which makes the magnetic moments at top and bottom surfaces antiparallel for $N_{\mathrm{slab}}$ even. For an odd number of layers ($N_{\mathrm{slab}}=9$ in Fig.\ref{fig:KVSO_slab_comparison_terminations}c), inversion symmetry $\mathcal{P}$ relates the top and bottom surfaces, enforcing identical dispersions and spin polarizations of the surface states.

\begin{figure}[t]
  \centering
     \includegraphics[width=\columnwidth]{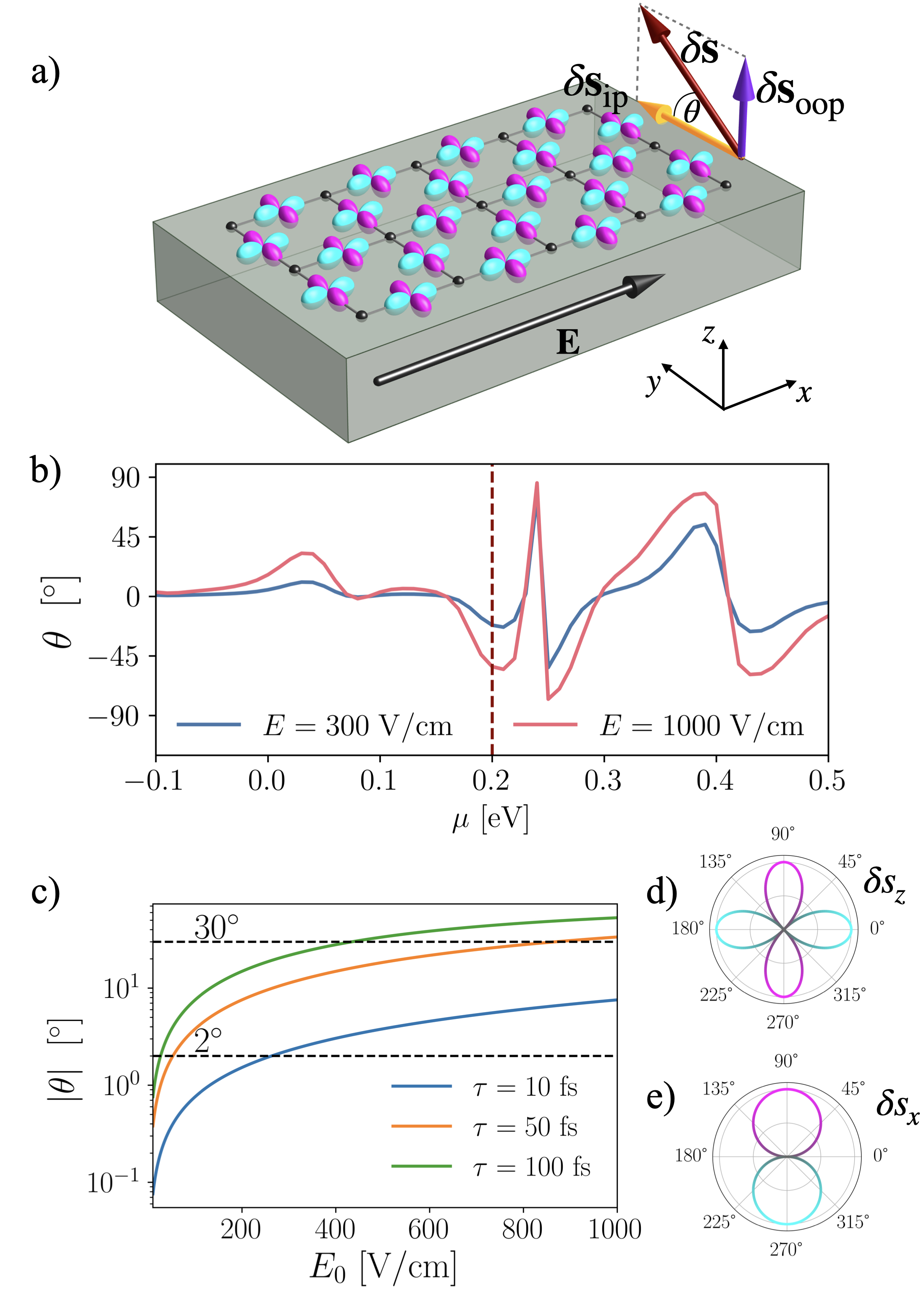}
    \caption{Comparison of in-plane (linear) and out-of-plane (nonlinear) contributions to the Edelstein response. (a) Schematic representing the altermagnetism at the VO-terminated surface $(001)$ of \KVSO{}. An electric field (black arrow) is applied along $\hat{\mathbf{x}}$,  the total response $\delta \mathbf{s}$ (red arrow) has in-plane and out-of-plane components denoted by $\delta \mathbf{s}_{\mathrm{ip}}$ (yellow arrow) and $\delta \mathbf{s}_{\mathrm{oop}}$ (red arrow), respectively. The angle $theta$ quantifies the strength of the nonlinear response with respect to the linear contribution. (b) Angle $\theta$ as function of the chemical potential, for two values of the applied field, using $\tau=100$ fs. (c) At fixed chemical potential $\mu=0.2$ eV, angle $\theta$ as function of the electric field, for several values of $\tau$, using $E_0=1000$ V/cm. Angular dependence of $\delta S_z$ (d) and $\delta S_x$ (e) with respect to in-plane electric field orientation. Radial axis in arbitrary units, and cyan and magenta color representing positive and negative sign, respectively. $\delta S_y$ (not shown) can be obtained from a 90$^{\circ}$ rotation of $\delta S_x$ plot. $N_{\mathrm{slab}}=7$ with VO termination was used in all calculations.}
    \label{fig:angle_spin}
\end{figure}

Figures~\ref{fig:KVSO_slab_comparison_terminations}d-f show the electronic structure of slabs with a SeK-terminated top surface. Since the bottom termination is VO in all cases, no differences are seen in the bottom-surface electronic structure. However, the projections onto the top surface show that the surface states around $\mu=0.2$ eV are no longer present. In this case, the uppermost VO layer is adjacent to SeK layers above and below, in contrast to the VO-terminated surface, in which there is no SeK layer above. This effectively reduces the local inversion asymmetry at the V sites. As a consequence, the electronic structure projected onto the top surface in Figs.~\ref{fig:KVSO_slab_comparison_terminations}e and f more closely resembles the bulk (see Figs.~\ref{fig:KVSO_slab_nlee}d and h).

Overall, slab parity controls the relative spin polarization of the top and bottom surfaces, whereas the surface termination affects both the character of the surface states and the nonlinear Edelstein effect. Figure~\ref{fig:nlee__vs_termination} shows a comparison of the responses for the two cases. In Fig.~\ref{fig:nlee__vs_termination}a and b, we show the surface nonlinear Edelstein response for VO- and SeK-terminated top surfaces, respectively. The peak at $\mu=0.2$~eV, present in the VO case, is absent for the SeK-terminated surface. This is expected because the large response is connected to the V-d surface states shown in Fig.~\ref{fig:KVSO_slab_comparison_terminations}c, which are not present for the SeK termination (see Fig.~\ref{fig:KVSO_slab_comparison_terminations}f). Figures~\ref{fig:nlee__vs_termination}c and d show the layer-resolved nonlinear Edelstein response in each case, demonstrating that the peak near the top surface, at $z\approx 60~\mathrm{\AA}$, is suppressed in the SeK-terminated case.

The large response at the surfaces with VO termination comes from an interplay of the time-reversal symmetry breaking associated with the surface altermagnetism and spin-orbit coupling. The response is enhanced by the local inversion symmetry breaking at the surface. Since the effect close to the Fermi level is dominated by V surface states, the presence of an extra layer of SeK atoms above the VO plane reduces the local inversion asymmetry of the uppermost VO layer. While a clean surface with VO termination will show a much larger response, the response is still finite for SeK termination.

The second-order surface Edelstein susceptibility reaches $\chi^{(2)}_{zxx}\approx125 \;\frac{ e^2 \tau^2 a^2}{(2\pi)^2 \hbar}$, which is two orders of magnitude larger than the previously reported value in Ref.~\cite{Trama2025}, in which a 2D model with Rashba SOC and altermagnetic exchange was adjusted to the band structure of RuO$_2$. The large value obtained in \KVSO{} is a consequence of the structure of the V-d surface states. In momentum space, the response is large close to avoided crossings of V-d states with opposite spin. These avoided crossings are generated by SOC, and in their vicinity the spin polarization acquires an in-plane component, which generates a finite value of the spin expectation value gradient $\frac{\partial s^{\alpha}_{n}(\mathbf{k})}{\partial k_i}$. This, combined with the large group velocity of the bands involved, that reaches approximately $0.5 \times 10^{6}$  m/s, generates a large integral in Eq.~\ref{eq:nlee_lr}.

\subsection{Comparison of first and second order Edelstein response}
\label{sec:nlee_comparison_lee}
So far we have discussed the nonlinear Edelstein response in \KVSO, focusing on the system size, and the surface termination. While the magnitude of $\chi^{(2)}_{zxx}$ in this system is large in comparison to related works \cite{Trama2025,Golub2025}, for a robust prediction it is vital to compare the effect with the linear contribution, which will always be present. The linear Edelstein effect is not related to surface altermagnetism, but to a combination of spin-orbit coupling and inversion symmetry breaking. As it was discussed previously, due to the symmetries of \KVSO, one expects a decoupling of both contributions (see Eq.~\ref{eq:edelstein_linear_and_nonlinear}), in which the linear and quadratic terms induce an in-plane and out-of-plane spin density, denoted by $\delta \mathbf{S}_{\mathrm{ip}}$ and $\delta \mathbf{S}_{\mathrm{oop}}$, respectively. The total spin density induced at the surface will have both components, as depicted in Fig.~\ref{fig:angle_spin}a. The angle $\theta$ measures the elevation of the total spin density with respect to the surface plane. Assuming an electric field pointing along the $a$-axis, $\mathbf{E}=E_0 \; \hat{\mathbf{x}}$, $\theta$ is given by:
\begin{equation}
    \tan \theta = \frac{|\delta \mathbf{S}_{\mathrm{oop}}|}{|\delta \mathbf{S}_{\mathrm{ip}}|} = \frac{\chi^{(2)}_{zxx}}{\chi^{(1)}_{zx}} E_0 \propto \tau E_0.
\end{equation}

 Figure \ref{fig:angle_spin}b shows the dependence of $\theta$ on the chemical potential $\mu$ for two values of $E_0$. At $\mu=0.2$ eV, where the surface states generate a large nonlinear response, $\theta$ reaches values of over $55^{\circ}$. At bit above in energy, at $\mu=0.23$~eV, $\theta$ reaches 90$^{\circ}$ because the linear response crosses zero. In Fig.~\ref{fig:angle_spin}c, we show the angle $\theta$ as function of the applied electric field, for several values of the relaxation time $\tau$. We note that electric fields of this magnitude are experimentally reachable and have been applied in the context of low-frequency nonlinear Hall measurements \cite{Yang2025b}. Additionally, relaxation times in the range of $\tau=$10-100 fs are reasonable for clean metallic samples at relatively low temperatures, and comparable to values used in existing literature \cite{Jiang2026, Li2024f}.

The nonlinear contribution is not only comparable, and in some cases larger, than its linear counterpart, but also has very different dependence on the orientation of the applied field. This is clear from Eq.~\eqref{eq:edelstein_linear_and_nonlinear}, and can also be seen in Fig.~\ref{fig:angle_spin}d and e, which show the angular dependence of $\delta S_z$ and $\delta S_x$ on the orientation of the electric field in the plane. $\delta S_y$ is not shown but it is related to $\delta S_x$ by a $90^{\circ}$ rotation. $\delta S_z$ shows two nodal directions along the diagonals, identical to the $d$-wave order parameter of the altermagnetic surface depicted in Fig.~\ref{fig:angle_spin}a. In contrast, the in-plane components have only one nodal direction each. In consequence, when reversing the direction of the applied field, the linear contribution changes sign, while the nonlinear one remains invariant. Additionally, if we reverse all magnetic moments, the nonlinear contribution should change sign, while the linear response, which is even under time reversal, will remain invariant. These symmetry considerations are relevant in the experimental detection. They allow, by changing the electric current orientation, to disentangle the surface nonlinear Edelstein signal from parasitic effects.

\section{Discussion and Outlook}
\label{sec::conclusions}

In this work we studied the magnetic system \KVSO{}, which has gained attention as a candidate for $d$-wave altermagnetism \cite{Jiang2025}. Ab initio studies predict a large splitting of more than 1 eV at the Fermi level \cite{Jiang2025,Jaeschke-Ubiergo2025}. The system is metallic and the large spin-splitting combined with its high critical temperature, make \KVSO{} a promising candidate for efficient charge-to-spin conversion. However, a more recent study determined, using neutron diffraction \cite{Sun2025a}, that the bulk of \KVSO{} orders antiferromagnetically. A suggestion to solve this apparent contradiction of the experimental evidence, was proposed in the context of surface altermagnetism \cite{Lange2026}. Here, using DFT, we studied the surface states of the electronic band structure of the altermagnetic and the antiferromagnetic phases of \KVSO. Notably, we show how the (001) surface of the antiferromagnetic bulk ordering, displays the same qualitative $d$-wave spin-splitting pattern as the bulk altermagnetic phase would show at the surface. This then strengthens the argument of \KVSO{} being an antiferromagnet in the bulk that shows $d$-wave surface altermagnetism in the (001) plane, which is what was measured by photoemision spectroscopy \cite{Jiang2025}, and very recently by spin-selective scanning tunneling microscopy \cite{Yang2026}.

By building a finite slab of \KVSO{}, we studied the surface electronic structure with different terminations and number of layers. We concluded that the most prominent surface states, with V-$d$ character, appear when the system is terminated at the VO layer, because there the V sites at the surface feel a stronger inversion symmetry breaking, in comparison with a SeK termination.

Additionally, we propose the nonlinear Edelstein effect \cite{Trama2025,Golub2025}, as a key experimental signature to probe the surface altermagnetism in \KVSO{}. We resolve the effect in the layers of a finite slab, showing that it cancels in the bulk, due to the antiferromagnetic ordering, and it peaks at the surface. The out-of-plane component of the induced spin density, has the same $d$-wave symmetry of the order parameter describing the surface altermagnetism. Moreover, the nonlinear contribution can be disentangled from the linear Edelstein response, which generates an in-plane spin density. For energies close to the Fermi level, the surface nonlinear Edelstein effect reaches 125 $\frac{ e^2 \tau^2 a^2}{(2\pi)^2 \hbar}$, which is two orders of magnitude larger than previous reports \cite{Trama2025,Golub2025}. This large response in \KVSO{} is attributed to avoided crossings of opposite-spin V-d surface states with large group velocity. While in our case SOC plays an important role in the response, it has been suggested that the nonlinear Edelstein effect can be finite even at zero SOC, where it would be constrained by spin symmetries \cite{Golub2025}. 

For realistic values of applied electric field, the second-order induced spin density can surpass the linear counterpart.
A possibility to measure the induced spin density, is to grow a non-magnetic metal on top of \KVSO{}. Application of a large enough electric field, will then induce an out-of-plane spin density at the interface, which will then diffuse into the metal, and could be detected by, for example, the inverse spin Hall effect. Rotating the applied electric field by 90$^{\circ}$, would reverse the sign of the signal. One of the challenges in the experimental detection of this effect is the roughness sensitivity. Terraces with opposite orientation of the magnetic moments will contribute with opposite signs to the total signal. Large magnetic domains and large terraces are important for avoiding cancellation.

This work contributes to the understanding of the highly discussed \KVSO{}, but it is also directly connected to a large family of altermagnetic candidates, whose structure resembles stacked layers of Lieb lattices \cite{Chang2025} such as \rbvto{}, \loas{} (with A=Fe,Mn), \vso{}, \ucr{}, \srcroas{}, among others \cite{Zhang2025a,Wei2025,lemoine2018,Ma2021,Jaeschke-Ubiergo2025,Campos2026}. Some of these systems have been predicted to be antiferromagnetic in the bulk, which makes them suitable candidates for surface altermagnetism, and the surface nonlinear Edelstein effect.

\textit{Acknowledgments} --- 
This work was funded by the German Research Foundation (DFG) through TRR 173-268565370 (Projects No.~A03 and B13), TRR 288-422213477 (Projects No.~A09 and B05), and Project No.~504261060 (Emmy Noether Programme). J. S. was supported partially by the DOE-Office of Science, DESC0026038. We acknowledge support by the Dynamics and Topology Center (TopDyn) funded by the State of Rhineland-Palatinate. This work was supported by the U.S. Department of Energy, Office of Science, under Award No. DE-SC0026038.
We acknowledge the high-performance computational facility of supercomputer “Mogon” at Johannes Gutenberg-Universität Mainz, Germany. 

\bibliography{references}
\bibliographystyle{apsrev4-2}

\newpage

\appendix
\section{Computational Details}
\label{sec:comp-details}
Our {\it ab initio} calculations were performed using density functional theory (DFT) within the plane-wave basis as implemented in the Vienna ab initio simulation package (VASP)~\cite{Kresse1996, Kresse1999}. For the exchange correlation functional, we used Perdew-Burke-Ernzerhof implementation of the generalized gradient approximation (GGA) optimized for solids (PBEsol), together with projector augmented-wave potentials~\cite{Blochl1994, Kresse1999}. All calculations were performed without Coulomb correlation, and both with and without spin-orbit coupling. When spin-orbit coupling is included, it is stated clearly in the text. In our DFT calculations we used a kinetic energy cutoff  of $10^{-7}$ eV for the plane-wave basis, and a $\Gamma$-centered $k$-point meshes of 12$\times$12$\times$7 (AM 1 uc) and 12$\times$12$\times$3 (AFM) $k$-point meshes for the momentum space calculations for the Brillouin zone integration of the bulk systems.
The surface spectral functions were calculated making use of Wannier functions, which we constructed using the VASP2WANNIER90~\cite{Marzari1997} and WANNIER90 codes~\cite{Mostofi2014}. The low-energy tight-binding Hamiltonian was defined in an effective Wannier basis including only K-p, V-d, and Se-p for and O-p orbitals, with all remaining degrees of freedom down-folded. Using the obtained tight-binding model, we calculate the surface spectral function for different terminations using the iterative Green’s function method, as implemented in the WannierTools package~\cite {Wu2017b}.

\newcommand{\ignorecolor}{gray!60}

\begin{table}[t]
\centering
\begin{tabular}{c c}
\hline\hline
Response & $\chi_{\alpha i}$ \\
\hline
$\chi^{(1)\; \mathrm{even}}$ &
$\displaystyle
\begin{pmatrix}
0 & \chi_{xy} & 0 \\
-\chi_{xy} & 0 & 0 \\
0 & 0 & 0
\end{pmatrix}
$
\\[1.5em]

$\chi^{(1)\;\mathrm{odd}}$ &
$\displaystyle
\begin{pmatrix}
 \chi_{xx} &0 & 0 \\
0 & - \chi_{xx} & 0 \\
0 & 0 & 0
\end{pmatrix}
$
\\
\hline\hline
\end{tabular}
\label{tab:linear_response}
\caption{Symmetry-constrained form of the linear Edelstein tensor under magnetic point group $4'm'm$.}
\end{table}

\begin{table*}[t]
\centering

\begin{tabular}{c c c c}
\hline\hline
Response &
$\chi_{xij}$ &
$\chi_{yij}$ &
$\chi_{zij}$ \\
\hline

$\chi^{(2)\;\mathrm{even}}_{\alpha ij}$ &
$\begin{pmatrix}
0 & 0 & {\color{\ignorecolor}0}\\
0 & 0 & {\color{\ignorecolor}\chi_{xyz}}\\
 {\color{\ignorecolor}0} & {\color{\ignorecolor}\chi_{xzy}} &  {\color{\ignorecolor}0}
\end{pmatrix}$ &
$\begin{pmatrix}
0 & 0 & {\color{\ignorecolor}-\chi_{xyz}}\\
0 & 0 &  {\color{\ignorecolor}0}\\
{\color{\ignorecolor}-\chi_{xzy}} &  {\color{\ignorecolor}0} &  {\color{\ignorecolor}0}
\end{pmatrix}$ &
$\begin{pmatrix}
0 & \chi_{zxy}&  {\color{\ignorecolor}0}\\
-\chi_{zxy} & 0 &  {\color{\ignorecolor}0}\\
 {\color{\ignorecolor}0} &  {\color{\ignorecolor}0} &  {\color{\ignorecolor}0}
\end{pmatrix}$ \\[2em]

$\chi^{(2)\; \mathrm{odd}}_{\alpha ij}$ &
$\begin{pmatrix}
0 & 0 & {\color{\ignorecolor}\chi_{xxz}}\\
0 & 0 &  {\color{\ignorecolor}0}\\
{\color{\ignorecolor}\chi_{xzx}} &  {\color{\ignorecolor}0} &  {\color{\ignorecolor}0}
\end{pmatrix}$ &
$\begin{pmatrix}
0 & 0 &  {\color{\ignorecolor}0}\\
0 & 0 & {\color{\ignorecolor}-\chi_{xxz}}\\
 {\color{\ignorecolor}0} & {\color{\ignorecolor}-\chi_{xzx}}  &  {\color{\ignorecolor}0}
\end{pmatrix}$ &
$\begin{pmatrix}
\chi_{zxx} & 0 &  {\color{\ignorecolor}0}\\
0 & -\chi_{zxx} &  {\color{\ignorecolor}0}\\
 {\color{\ignorecolor}0} &  {\color{\ignorecolor}0} &  {\color{\ignorecolor}0}
\end{pmatrix}$ \\

\hline\hline
\end{tabular}
\label{tab:nonlinear_response}
\caption{Symmetry-constrained form of the second-order Edelstein tensor under magnetic point group $4'm'm$.}
\end{table*}

\section{Additional details on the nonlinear Edelstein response}

In the Boltzmann formalism, the total spin per unit cell can be written in terms of the electron distribution as
\begin{equation}
\mathbf{s} = \frac{a^2}{(2\pi)^2}\sum_n \int d\mathbf{k} \;\mathbf{s}_n(\mathbf{k}) f_{n}(\mathbf{k})\;,
\end{equation}

with $a$ denoting the lattice constant, $n$ the band index, $\mathbf{s}_n(\mathbf{k})$ the spin expectation value, and $f_{n}(\mathbf{k})=f(\varepsilon_n(\mathbf{k}))$ the carrier distribution evaluated at the energy $\varepsilon_n(\mathbf{k})$. Under the application of a constant and homogeneous electric field $\mathbf{E}$, the electron distribution is modified. Using the constant relaxation time approximation we can write

\begin{equation}
\frac{e}{\hbar}\,\mathbf{E}\cdot\nabla_{\mathbf{k}} f_{n}(\mathbf{k})
=
\frac{f_{n}(\mathbf{k})-f_{n}^0(\mathbf{k})}{\tau},
\end{equation}
with $\tau$ denoting the relaxation time and $f^{0}_{n}(\mathbf{k}) = f_{0}(\varepsilon_n(\mathbf{k}))$, denoting the Fermi-Dirac distribution evaluated at the energy $\varepsilon_n(\mathbf{k})$, of a state with band index $n$ and momentum $\mathbf{k}$.
Recursive replacement of the distribution allows us to expand it up to second order in the field as
\begin{equation}
f_{n}(\mathbf{k}) = f^{0}_{n}(\mathbf{k}) + \frac{e\tau}{\hbar} \mathbf{E}\cdot \nabla f^0_{n}(\mathbf{k}) + \frac{e^2 \tau^2}{\hbar^2} \frac{\partial^2 f^0_n(\mathbf{k})}{\partial k_i \partial k_j} E_i E_j.
\end{equation}

We can then directly write the second-order Edelstein susceptibility as:
\begin{equation}
\chi^{(2)}_{\alpha ij} = \frac{e^2 \tau^2 a^2}{(2\pi)^2 \hbar^2} \ \sum_n \int d\mathbf{k} \; s^{\alpha}_{n}(\mathbf{k}) \frac{\partial^2 f^0_n(\mathbf{k})}{\partial k_i \partial k_j}.
\end{equation}
Integration by parts, and substitution of $\frac{\partial f^0_n(\mathbf(k))}{\partial k_j} = \frac{\partial \varepsilon_n(\mathbf{k})}{\partial k_j} \frac{\partial f^0_n(\mathbf{k})}{\partial \varepsilon}$ leads to
\begin{equation}
\chi^{(2)}_{\alpha ij} = -\frac{e^2 \tau^2 a^2}{(2\pi)^2 \hbar} \sum_n \int d\mathbf{k} \; \frac{\partial s^{\alpha}_n(\mathbf{k})}{\partial k_i} v^{j}_n(\mathbf{k}) \frac{\partial f^0_n(\mathbf{k})}{\partial \varepsilon}.
\end{equation}

At zero temperature we take $\frac{\partial f^0_n(\mathbf{k})}{\partial \varepsilon} = -\delta(\varepsilon_n(\mathbf{k})-E_F)$, and one recovers the expression of Eq.~\eqref{chi_2}. In our calculations, we use a small but finite temperature of $k_B T=5$ meV ($T\approx 58$ K). We note that in our units system, the susceptibility $\chi^{(2)}_{\alpha ij}$ relates an applied electric field, with an induced spin per unit cell. In order to get units of magnetization, one needs to add an extra factor of $-\frac{g\mu_B}{\hbar}$.

To calculate the partial derivative of the spin expectation value, for faster convergence of the integral, we make use of the identity:
\begin{equation}
\frac{\partial s^{\alpha}_n (\mathbf{k})}{\partial k_i} = 2 \Re \sum_{m\neq n} \frac{\bra{\psi_{n\mathbf{k}}} \frac{\partial H(\mathbf{k})}{\partial k_i} \ket{\psi_{m\mathbf{k}}}\bra{\psi_{m\mathbf{k}}} \hat{S}\ket{\psi_{n\mathbf{k}}}}{\varepsilon_n(\mathbf{k})-\varepsilon_m(\mathbf{k})},
\end{equation}
which turns the expression into an interband summation of the spin operator $\hat{S}^{\alpha}$ and the velocity operator $\hat{v}^{i}=\frac{1}{\hbar}\frac{\partial H(\mathbf{k})}{\partial k_i}$. To take care of the denominator at the degeneracies we introduce a finite broadening $\eta$ by replacing $\varepsilon_{n}(\mathbf{k})\rightarrow \varepsilon_{n}(\mathbf{k}) + i \eta$, to finally obtain
\begin{equation}
\frac{\partial s^{\alpha}_n (\mathbf{k})}{\partial k_i} = 2 \hbar \Re \sum_{m}\frac{\varepsilon_n(\mathbf{k})-\varepsilon_m(\mathbf{k}) }{(\varepsilon_n(\mathbf{k})-\varepsilon_m(\mathbf{k}))^2 + \eta^2} s^{\alpha}_{mn}(\mathbf{k}) v^{i}_{nm}(\mathbf{k}).
\end{equation}
Here, $s^{\alpha}_{mn}(\mathbf{k})$ and $ v^{i}_{nm}(\mathbf{k})$ are the matrix elements of the spin operator $\hat{S}^{\alpha}$ and the velocity operator $\hat{v}^{i}$, respectively. In our calculations, we set $\eta=1$ meV.

\section{Symmetry constrained form of the Edelstein response}
\label{symmetry_edelstein}

The surface $(001)$ of \KVSO\ has magnetic point group $4'm'm$, with generators $C_{4z}\mathcal{T}$, $m_{100}\mathcal{T}$, $m_{010}\mathcal{T}$, $m_{110}$, $m_{1\bar{1}0}$, where $\mathcal{T}$ denotes time reversal, $C_{4z}$ is a four-fold rotation around the out-of-plane axis, and $m_v$ represents a mirror normal to direction $v$.

The Edelstein tensors can be decomposed into parts that are even and odd with respect to time reversal symmetry. In particular we have
\begin{equation}
    \chi^{(1)}_{\alpha i} = \chi^{(1)\; \rm even }_{\alpha i} + \chi^{(1)\; \rm odd }_{\alpha i}
\end{equation}
and
\begin{equation}
    \chi^{(2)}_{\alpha ij} = \chi^{(2)\; \rm even }_{\alpha ij} + \chi^{(2)\; \rm odd }_{\alpha ij},
\end{equation}
for the linear and nonlinear Edelstein tensors, respectively. When neglecting interband transitions and considering only the intraband contribution, which is expected to dominate in a metallic system such as \KVSO, the corresponding response tensors have well defined parity under time reversal. In particular $\chi^{(1)}_{\alpha ij} \approx \chi^{(1)\; \mathrm{even}}_{\alpha ij}$ and $\chi^{(2)}_{\alpha ij} \approx \chi^{(2)\; \mathrm{odd}}_{\alpha ij}$.

In Tables I and  II we show the symmetry constrained form of the full tensors (both time-reversal even and odd contributions) under the surface magnetic group of \KVSO. Since the surface is two-dimensional, only the spatial $x$ and $y$ direcitons are relevant. Spatial $z$ components are written in gray to distinguish them from the relevant $x$ and $y$ components (written in black). From this symmetry constrained form one can recover the spin density induced by an in-plane electric field of Eq.~\eqref{eq:edelstein_linear_and_nonlinear}.

\end{document}